\documentclass[runningheads]{llncs}
\usepackage[T1]{fontenc}
\usepackage{graphicx}
\usepackage{orcidlink}
\usepackage{hyperref}
\begin{document}
\title{A Fuzzy Supervisor Agent Design for Clinical Reasoning Assistance in a Multi-Agent Educational Clinical Scenario Simulation}
\titlerunning{A FSA Design for CR Assistance in a MAECSS}
%
\author{Weibing Zheng\orcidlink{0000-0002-6496-1906}  \and Laurah Turner\orcidlink{0000-0002-4567-1313} 
\and Jess Kropczynski\orcidlink{0000-0002-7458-6003} \and Murat Ozer \and Seth Overla \and Shane Halse\orcidlink{0000-0002-0799-6715}}
\authorrunning{Zheng, W et al.}
%
\institute{University of Cincinnati, Cincinnati, OH, USA  
\email{zhengwb@mail.uc.edu}}
\maketitle   
\begin{abstract}
Assisting medical students with clinical reasoning (CR) during clinical scenario training remains a persistent challenge in medical education. This paper presents the design and architecture of the Fuzzy Supervisor Agent (FSA), a novel component for the Multi-Agent Educational Clinical Scenario Simulation (MAECSS) platform. The FSA leverages a Fuzzy Inference System (FIS) to continuously interpret student interactions with specialized clinical agents (e.g., patient, physical exam, diagnostic, intervention) using pre-defined fuzzy rule bases for professionalism, medical relevance, ethical behavior, and contextual distraction. By analyzing student decision-making processes in real-time, the FSA is designed to deliver adaptive, context-aware feedback and provides assistance precisely when students encounter difficulties. This work focuses on the technical framework and rationale of the FSA, highlighting its potential to provide scalable, flexible, and human-like supervision in simulation-based medical education. Future work will include empirical evaluation and integration into broader educational settings. More detailed design and implementation is~\href{https://github.com/2sigmaEdTech/MAS/}{open sourced here}.


\end{abstract}

\section{Introduction}
Training in clinical reasoning skills is an essential component of medical education; however, providing scalable, real-time, and precise guidance to students during simulated patient encounters remains a persistent challenge. Traditional methods of assessment and assistance often rely on post-hoc evaluations or limited availability of medical educators. With the increasing complexity of clinical scenarios, there is a pressing need for innovative solutions that can provide immediate context-based assistance to medical learners.

Current simulation-based medical education includes human-led debriefings, rule-based virtual patient systems, and emerging AI-powered platforms. While human educators offer nuanced feedback, their participant is resource-intensive and difficult to scale for large learners~\cite{issenbergFeaturesUsesHighfidelity2005}. While rule-based simulators can automate some aspects of feedback but often lack the flexibility to interpret the subtle, context-dependent nature of student performance, often leading to rigid or overly simplistic guidance~\cite{cookTechnologyenhancedSimulationHealth2011}. Recent advances in Artificial Intelligence (AI) and Large Language Models (LLMs) have led to the development of AI-powered medical simulation platforms, such as 2-Sigma, which has been pivoted at College of Medicine, University of Cincinnati~\cite{zhengUserCenteredIterativeDesign2025}. However, a post-use survey indicates that most medical students express a strong need for real-time and precise assistance during clinical scenario simulation.

\begin{figure}
\centering
\includegraphics[width=0.99\textwidth]{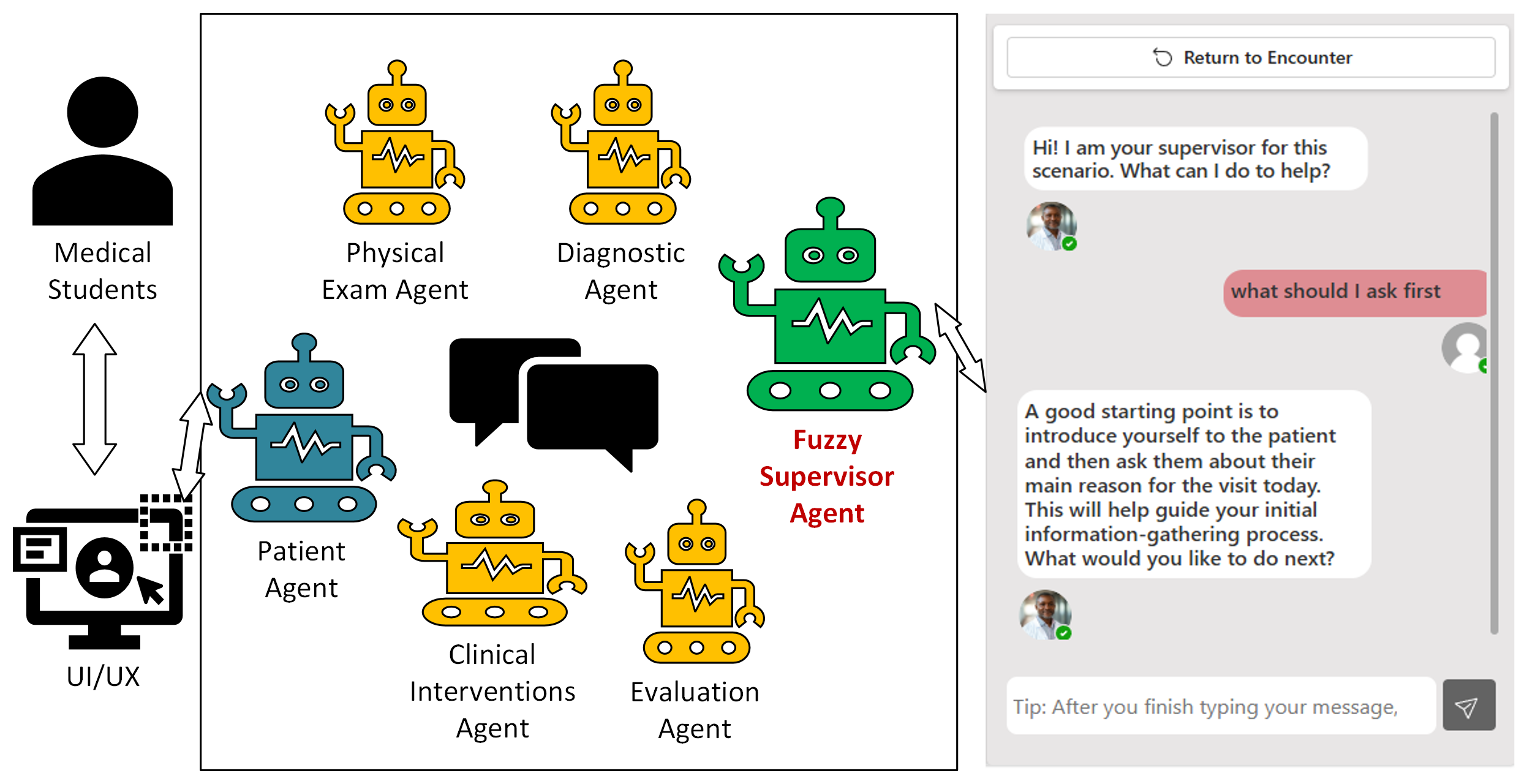}
\caption{Fuzzy Supervisor Agent within Multi-Agent Clinical Scenario Simulation}\label{fig1:fuzzysupervisor}
\end{figure}

To address these needs, we propose the Fuzzy Supervisor Agent (FSA), a novel and intelligent agent designed to interpret the ambiguous and variable actions of medical students within a Multi-Agent Educational Clinical Scenario Simulation (MAECSS), as illustrated in Figure~\ref{fig1:fuzzysupervisor}. Using fuzzy logic, the agent continuously monitors medical student interactions with simulated patient, physical exam, diagnostic, and clinical intervention agents, providing adaptive real-time feedback that reflects the complexity of authentic clinical reasoning from professionalism, medical relevance, ethical behavior, and contextual distraction~\cite{zhengLLMasaFuzzyJudgeFineTuningLarge2025}. The FSA bridges the gap between rigid automation and medical educator's expertise and meets the needs of scale, real-time, and precise clinical assistance. This paper details the design and rationale of the FSA, demonstrating a novel approach to automated, context-aware student assistance on the MAECSS platform. Future work will focus on empirical evaluation and integration into broader educational settings.

\section{System Design and Methodology}
\subsection{Multi-Agent Educational Clinical Scenario Simulation (MAECSS)}
\begin{figure}
\centering
\includegraphics[width=0.99\textwidth]{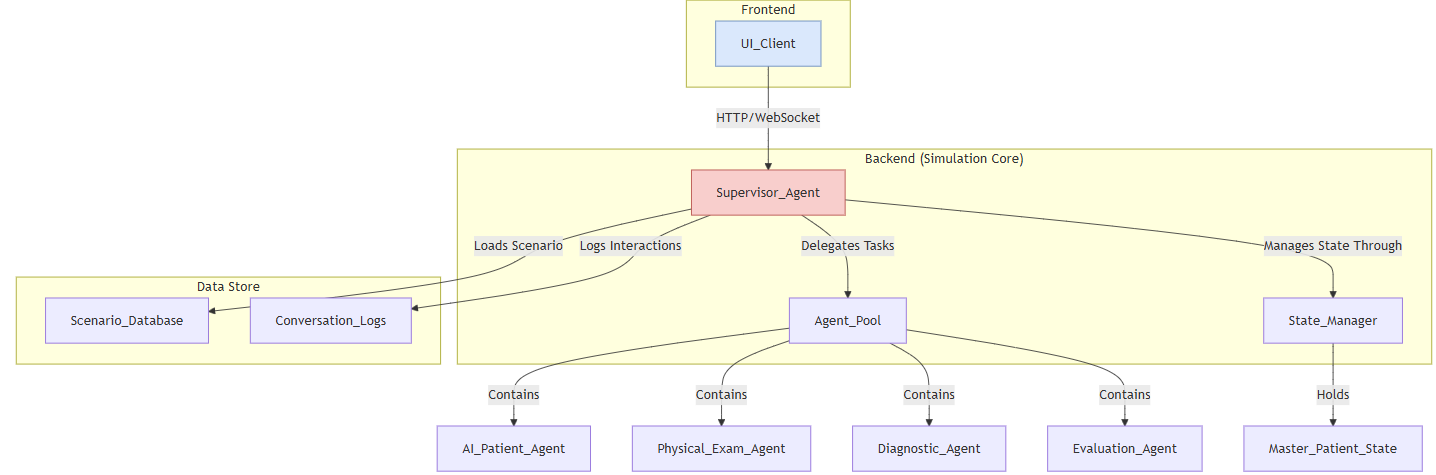}
\caption{Multi-Agent Clinical Scenario Simulation High Level Architecture}\label{fig2:MASSArchitecture}
\end{figure}

As shown in Figure~\ref{fig1:fuzzysupervisor}, MAECSS platform is designed to mimic authentic doctor-patient conversation encounters for medical students~\cite{zhengUserCenteredIterativeDesign2025}. The high-level architecture is shown in Figure~\ref{fig2:MASSArchitecture}, and more detailed design and implementation is~\href{https://github.com/2sigmaEdTech/MAS/}{open sourced}. Medical students log in the platform UI and choose the simulated clinical case to start. The platform is orchestrated by the central FSA with several autonomous clinical agents to perform different roles to help medical students during the diagnostic process. Each agent represents a distinct aspect of the MAECSS workflow: the Patient Agent simulates patient responses and symptoms; the Physical Exam Agent manages examination findings; the Diagnostic Agent handles test ordering and results; the Clinical Intervention Agent oversees therapeutic actions; the Evaluation Agent assesses overall student performance and generates the final report; the FSA as an orchestrator to manage all other agents and provides real-time feedback and guidance. Medical students interact with these agents through a unified UI, generating a rich interaction and behavioral data, such as questions asked, tests ordered, and interventions performed, which serve as the primary input for the FSA.

\subsection{Fuzzy Supervisor Agent (FSA) Architecture}
\begin{figure}
\centering
\includegraphics[width=0.99\textwidth]{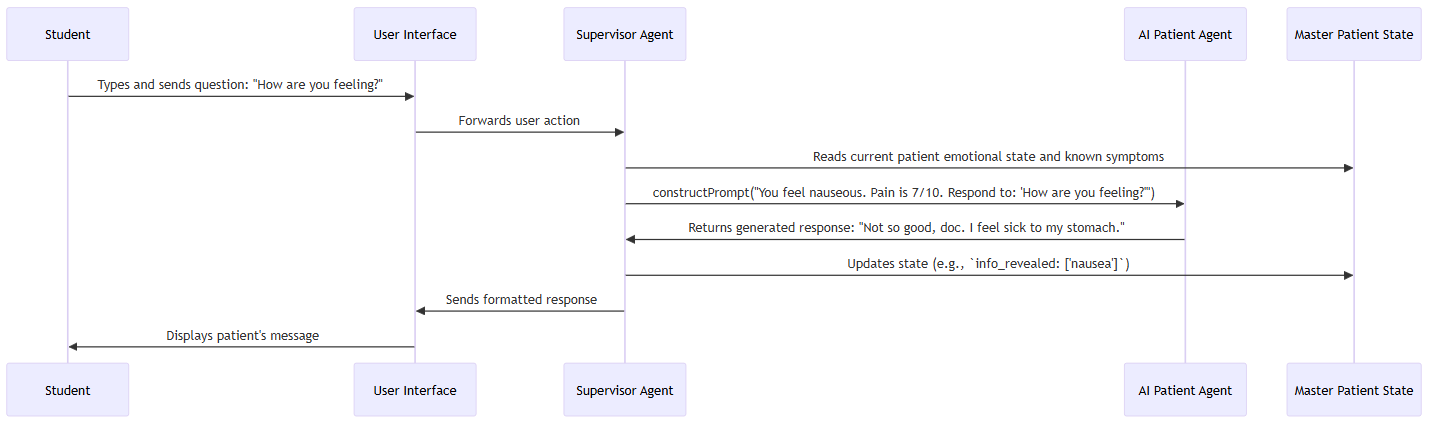}
\caption{A FSA Centered Sequence Diagram Example}\label{fig3:MASSSequenceDiagram}
\end{figure}

The FSA, which functions as a gatekeeper and orchestrator, continuously monitors student interactions with all clinical agents in simulation scenarios. Figure~\ref{fig3:MASSSequenceDiagram} illustrates a sequence diagram of FSA centered on the interactions of medical students, UI, FSA, and clinical agents. It also tracks a variety of data points, including the relevance and appropriateness of the questions asked, the correctness and sequence of physical exams performed, the selection and timing of diagnostic tests, and ethical considerations in patient management. Also, the agent records temporal metrics such as the time elapsed between actions and the frequency of off-task behaviors (e.g., repeated irrelevant questions). As shown in Figure ~\ref{fig1:fuzzysupervisor}, these data points and the conversation log are stored in different data tables.  This comprehensive monitoring enables the agent to construct a nuanced, real-time profile of student performance and to provide precise personalized assistance.

\subsection{The Fuzzy Inference System (FIS) Design}
Beyond the orchestrator role for the MAECSS, the FSA contains an FIS that interprets the collected data to assess student performance and determine the appropriate level of assistance. The FIS uses several fuzzy criteria with fuzzy sets that reflect the ambiguous and uncertain nature of clinical reasoning, identified by medical experts, and fine-tuned large language models through iterative feedback~\cite{zhengLLMasaFuzzyJudgeFineTuningLarge2025}. The FIS is designed to assess student actions in real time, providing precise adaptive assistance that reflects the complexity of clinical decision making. The FIS consists of three main components: input fuzzy criteria, fuzzy sets with different levels, and a fuzzy rule base.

\begin{itemize}
    \item \textbf{Professionalism} refers to maintaining appropriate conduct and respecting boundaries: \textbf{1. Unprofessional, 2. Borderline, 3. Appropriate}.
    \item \textbf{Medical Relevance} tracks whether the students remain focused on the case: \textbf{1. Irrelevant, 2. Partially relevant, 3. Relevant}.
    \item \textbf{Ethical Behavior} ensures actions prioritize patient safety and consent: \textbf{1. Dangerous, 2. Unsafe, 3. Questionable, 4. Mostly safe, 5. Safe}.
    \item \textbf{Contextual Distraction} assesses whether each message is relevant to the ongoing conversation: \textbf{1. Highly distracting, 2. Moderately distracting, 3. Questionable, 4. Not distracting}.
\end{itemize}

\subsubsection{Fuzzy Rule Base}
The FIS utilizes a comprehensive fuzzy rule base to interpret student actions and determine the appropriate level of assistance as Table~\ref{tab:fuzzy_rules}.

\begin{table}[ht]
\centering
\caption{Illustrative Fuzzy Rules for Assistance Level Determination}
\label{tab:fuzzy_rules}
\begin{tabular}{|p{10.2cm}|p{1.7cm}|}
\hline
\textbf{Condition(s)} & \textbf{Assistance} \\
\hline
Professionalism is Unprofessional \textbf{OR} Ethical Behavior is Dangerous & Very High \\
Medical Relevance is Irrelevant \textbf{AND} Contextual Distraction is Highly distracting & Very High \\
Professionalism is Borderline \textbf{AND} Ethical Behavior is Unsafe & High \\
Medical Relevance is Partially relevant \textbf{AND} Contextual Distraction is Moderately distracting & High \\
Professionalism is Appropriate \textbf{AND} Medical Relevance is Relevant \textbf{AND} Ethical Behavior is Safe \textbf{AND} Contextual Distraction is Not distracting & Low \\
Professionalism is Appropriate \textbf{AND} Medical Relevance is Relevant \textbf{AND} Ethical Behavior is Mostly safe \textbf{AND} Contextual Distraction is Questionable & Medium \\
Medical Relevance is Irrelevant \textbf{AND} Ethical Behavior is Questionable & High \\
Professionalism is Borderline \textbf{AND} Medical Relevance is Partially relevant \textbf{AND} Contextual Distraction is Moderately distracting & Medium \\
Ethical Behavior is Unsafe \textbf{OR} Dangerous & Very High \\
Contextual Distraction is Highly distracting & High \\
Professionalism is Appropriate \textbf{AND} Medical Relevance is Relevant \textbf{AND} Ethical Behavior is Safe \textbf{AND} Contextual Distraction is Not distracting & Minimal \\
Any criterion is at its lowest level (e.g., Unprofessional, Irrelevant, Dangerous, Highly distracting) & Highest \\
\hline
\end{tabular}
\end{table}

These rules enable the FSA to provide nuanced, context-aware feedback, prioritizing urgent intervention when student actions are unsafe or unprofessional, and minimizing assistance when performance is appropriate and safe.

\subsection{Real-Time Assistance Mechanism}
When the FSA detects the assistance level is High or above, it will be ready to deliver context-sensitive real-time feedback to medical students. This feedback, as shown in Figure~\ref{fig1:fuzzysupervisor}, shows hints suggesting more relevant questions, clarifying prompts to reconsider an ethical issue, or notifications highlighting areas for improvement. The FSA is minimally intrusive, providing assistance only when necessary to support learning while preserving student autonomy.

\section{Design Rationale and Example Use Case}
The design of the FSA is grounded in the need for scalable, real-time, and precise assistance in MAECSS. Fuzzy logic enables the system to interpret ambiguous or partially correct student actions, reflecting the complexity of real-world clinical reasoning. The input variables —professionalism, medical relevance, ethical behavior, and contextual distraction — were selected based on expert consensus and iterative review, as these dimensions capture the most common areas where students struggle during clinical scenario simulation encounters. 

Here is an example that shows the FSA continuously analyzes student actions and provides targeted, context-aware assistance to facilitate clinical reasoning development in MEACSS. Consider a scenario where a medical student manages a simulated patient with chest pain. If the student asks several questions that are only partially relevant to the chief complaint, the FSA evaluates the medical relevance as "Partially relevant" and contextual distraction as "Moderately distracting." Based on the fuzzy rule base, this results in a "High" assistance level, prompting the system to deliver real-time guidance such as: \emph{"Consider focusing your questions on symptoms related to chest pain and cardiovascular risk factors."} Later, if the student attempts an intervention without obtaining patient consent, the ethical behavior score drops to "Unsafe." The FSA then escalates to a "Very High" assistance level and issues a prompt: \emph{"Before proceeding, ensure you have explained the procedure and obtained the patient's consent."}

\section{Conclusion and Future Work}
This paper presents the design and rationale of a FSA for clinical reasoning assistance within a MAECSS platform. Using fuzzy logic, the FSA is capable of interpreting partially correct and ambiguous student actions, enabling real-time, context-aware feedback across key domains such as professionalism, medical relevance, ethical behavior, and contextual distraction. The agent architecture is designed to provide scalable, adaptive, and human-like supervision, addressing the limitations of both rule-based systems and resource-intensive human feedback in simulation-based medical education. It lays the foundation for the following implementation and empirical studies. Next steps will include implementing the FSA in MEACSS and further refining the FSA to support intelligent clinical reasoning assistance.

\begin{credits}
\subsubsection{\ackname} The study was supported by 2-Sigma and ARC at the University of Cincinnati. The manuscript and GitHub code used Copilot to polish and help. We thank all medical experts for fuzzy criteria/sets/rules and reviewers for their feedback.


\end{credits}
%
%
%
\bibliographystyle{splncs04}
\bibliography{references}

\begin{thebibliography}{1}
\providecommand{\url}[1]{\texttt{#1}}
\providecommand{\urlprefix}{URL }
\providecommand{\doi}[1]{https://doi.org/#1}

\bibitem{cookTechnologyenhancedSimulationHealth2011}
Cook, D.A., Hatala, R., Brydges, R., Zendejas, B., Szostek, J.H., Wang, A.T.,
  Erwin, P.J., Hamstra, S.J.: Technology-enhanced simulation for health
  professions education: A systematic review and meta-analysis. JAMA
  \textbf{306}(9),  978--988 (Sep 2011). \doi{10.1001/jama.2011.1234}

\bibitem{issenbergFeaturesUsesHighfidelity2005}
Issenberg, S.B., McGaghie, W.C., Petrusa, E.R., Lee~Gordon, D., Scalese, R.J.:
  Features and uses of high-fidelity medical simulations that lead to effective
  learning: A {{BEME}} systematic review. Medical Teacher  \textbf{27}(1),
  10--28 (Jan 2005). \doi{10.1080/01421590500046924}

\bibitem{zhengUserCenteredIterativeDesign2025}
Zheng, W., Turner, L., Kropczynski, J., Ozer, M., Kelleher, M., Overla, S.,
  Halse, S.: The {{User-Centered Iterative Design}} of an {{LLM-Powered
  Educational Scenario Simulator}} for {{Clinical Reasoning}}. CS \& IT
  Conference Proceedings  \textbf{15}(10) (May 2025),
  \url{http://csitcp.com/abstract/15/1510csit19}

\bibitem{zhengLLMasaFuzzyJudgeFineTuningLarge2025}
Zheng, W., Turner, L., Kropczynski, J., Ozer, M., Nguyen, T., Halse, S.:
  {{LLM-as-a-Fuzzy-Judge}}: {{Fine-Tuning Large Language Models}} as a
  {{Clinical Evaluation Judge}} with {{Fuzzy Logic}} (Jun 2025).
  \doi{10.48550/arXiv.2506.11221}, \url{http://arxiv.org/abs/2506.11221}

\end{thebibliography}

\end{document}